\documentclass[twocolumn,showpacs,preprintnumbers,amsmath,amssymb,nofootinbib]{revtex4-1}
\usepackage{here}
\usepackage{comment,braket}
\usepackage{epsf}
\usepackage{amsmath}
\usepackage{graphics}
\usepackage{amsfonts}
\usepackage{amssymb}
\usepackage{latexsym}
\usepackage{color}
\usepackage{ulem}
\input{colordvi.tex}
\usepackage{feynmp}
\usepackage{natbib}
\usepackage[pdftex]{graphicx}
\usepackage{bm}
\usepackage[pdftex]{hyperref}
\usepackage[svgnames]{xcolor}
\definecolor{phthaloblue}{rgb}{0.0, 0.06, 0.54}
\hypersetup{
    colorlinks=true,
    linkcolor=phthaloblue,
    citecolor=blue,
    filecolor=blue,
    urlcolor=phthaloblue,
    }
    
\newcommand{\beq}{\begin{eqnarray}} 
\newcommand{\eeq}{\end{eqnarray}}

\def\({\left(}
\def\){\right)}
\def\[{\left[}
\def\]{\right]}

\def\nn{\nonumber \\}

\def\Tr{{\rm Tr}}

\def\lmk{\left(}
\def\rmk{\right)}
\def\lkk{\left[}
\def\rkk{\right]}

\def\la{\left<}
\def\ra{\right>}
\newcommand{\eq}[1]{Eq.~(\ref{#1})}
\newcommand{\bel}[1] {\begin{equation}\label{#1}}
\newcommand{\beal}[1] {\begin{eqnarray}\label{#1}}
\newcommand{\be}{\begin{equation}}
\newcommand{\ee}{\end{equation}}
\newcommand{\bea}{\begin{array}} 
\newcommand{\eea}{\end{array}}

\def\del{\partial}
\def\GeV{\ {\rm GeV}}

\begin{document}

\title{Vacuum Decay in Real Time and Imaginary Time Formalisms}

\author{Mark P.~Hertzberg$^{1,2}$}
\email{mark.hertzberg@tufts.edu}
\author{Masaki Yamada$^1$}
\email{masaki.yamada@tufts.edu}
\affiliation{$^1$Institute of Cosmology, Dept.~of Physics and Astronomy, Tufts University, Medford, MA 02155, USA\looseness=-1}
\affiliation{$^2$Department of Physics, Tokyo Institute of Technology, Ookayama, Meguro-ku, Tokyo 152-8551, Japan}

\date{\today}

\begin{abstract}
We analyze vacuum tunneling in quantum field theory in a general formalism by using the Wigner representation. In the standard instanton formalism, one usually approximates the initial false vacuum state by an eigenstate of the field operator, imposes Dirichlet boundary conditions on the initial field value, and evolves in imaginary time. This approach does not have an obvious physical interpretation. However, an alternative approach does have a physical interpretation: in quantum field theory, tunneling can happen via classical dynamics, seeded by initial quantum fluctuations in both the field and its momentum conjugate, which was recently implemented in Ref.~\cite{Braden:2018tky}. We show that the Wigner representation is a useful framework to calculate and understand the relationship between these two approaches. We find there are two, related, saddle point approximations for the path integral of the tunneling process: one corresponds to the instanton solution in imaginary time and the other one corresponds to classical dynamics from initial quantum fluctuations in real time. The classical approximation for the dynamics of the latter process is justified only in a system with many degrees of freedom, as can appear in field theory due to high occupancy of nucleated bubbles, while it is not justified in single particle quantum mechanics, as we explain. We mention possible applications of the real time formalism, including tunneling when the instanton vanishes, or when the imaginary time contour deformation is not possible, which may occur in cosmological settings.
\end{abstract}

\maketitle

\section{Introduction}

The subject of quantum mechanical tunneling is an essential topic in modern physics, with a range of applications, including nuclear fusion \cite{Balantekin:1997yh}, diodes \cite{1958PhRv..109..603E}, atomic physics \cite{1982PhRvL..49...57B}, quantum field theory \cite{1977PhRvD..15.2922F}, cosmological inflation \cite{Guth:1980zm}, etc. In the context of a possible landscape of classically stable vacua in field theory, motivated by considerations in string theory \cite{Douglas:2003um}, it is essential to determine the quantum tunneling rate from one vacuum to the next. This has ramifications for the stability of our current electroweak vacuum \cite{Sher:1993mf}, as well as for the viability of inflationary models \cite{Guth:2012ww}, and may have ramifications for the cosmological constant problem \cite{Abbott:1984qf}. 

In ordinary non-relativistic quantum mechanics of a single particle, quantum tunneling can be calculated in principle by a direct solution of the time dependent Schr\"odinger equation. However, our interest here is that of quantum field theory. In this case, a direct solution of the Schr\"odinger equation is notoriously difficult, and so approximation schemes are needed. The most famous approximation method, which is analogous to the WKB approximation in non-relativistic quantum mechanics, involves the computation of the Euclidean instanton solution from one vacuum to another \cite{Coleman:1977py,Callan:1977pt}. This leads to the well known estimate for the decay rate per unit volume $\Gamma\propto e^{-S_E}$, where $S_E$ is the bounce  action of a solution of the classical equations of motion in imaginary time. This method is generally thought to be accurate when the bounce action $S_E$ is large; which evidently corresponds to exponentially suppressed decay rates.

Since the above method involves non-intuitive features, namely a restriction to Dirichlet boundary conditions on the field and dynamics in imaginary time, it begs the question whether there may be other formulations of the tunneling process. Furthermore, if one moves to more general settings, such as in cosmology, there may not always be the usual instanton solution, so one wonders whether other formulations can be employed instead. In this paper we will investigate under what circumstances an alternative approach to tunneling, from classical evolution of fields whose initial conditions are drawn from some approximation to the initial wave-function, can provide an alternative formulation for decay.%
\footnote
{See Refs.~\cite{Turok:2013dfa, Bramberger:2016yog} for a different approach using complex classical trajectories.
}

This work was motivated by the very interesting work of Ref.~\cite{Braden:2018tky}. In that work they numerically obtained a tunneling rate from a false vacuum in $1+1$ dimensional spacetime by solving for the classical dynamics of a scalar field starting from initial conditions generated by a Gaussian distribution. The method was to consider many realizations of initial conditions and then to calculate the ensemble-averaged tunneling rate. For their choice of parameters they found that the tunneling rate was similar to the one calculated by the instanton method.  

This leads to several natural questions: (i) what is the relationship between these two approaches? One is in an imaginary time formalism, the other is in a real time formalism; so how, if at all, are they related? (ii) Under what circumstances are the rates comparable to each other? (iii) Under what circumstances are these approaches valid? It is known that the instanton method requires the bounce action to be large to justify a semi-classical approximation, but what is the corresponding statement for the other real time method? 

In this paper, we address these questions. We will argue that this real time analysis from classical dynamics is not identical to, but is very closely related to, the instanton tunneling process. We will show that for simple choices of parameters, the two rates are parametrically similar. However, things are more complicated for potentials with unusual features, which we will discuss, and there can be advantages to the real time formulation in special circumstances. We will make use of the Wigner representation as it will provide a general formalism to cleanly identify these two complementary approaches. We will discuss under what circumstances the classical dynamics is justified, explain why this would fail in single particle quantum mechanics, and discuss some cosmological applications.

Our paper is organized as follows: 
In Section \ref{Imaginary} we recap the standard instanton contribution to the decay.
In Section \ref{Real} we present a more general formalism, using the Wigner representation, which allows us to describe these two approaches within a single framework.
In Section \ref{Condition} we discuss the conditions under which the classical dynamics method is applicable.
In Section \ref{Tunneling} we estimate and compare the tunneling rates.
Finally, in Section \ref{Discussion} we discuss our findings.

\section{Standard Euclidean Formalism}\label{Imaginary}

Let us begin by recapping the standard approach to vacuum decay, which occurs within the confines of a Euclidean, or imaginary time, formalism. In this approach the decay rate can be calculated from the imaginary part of the vacuum energy $E_0$ as 
\be
 \Gamma_I = - 2\, {\rm Im} E_0, 
 \ee
 where
 \be
 E_0 = - {\rm lim}_{{\cal T} \to \infty} \frac{\ln Z}{{\cal T}}, 
\ee
and $Z$ is defined by 
\beq
 Z \equiv \left< \phi_i \right\vert e^{-H {\cal T}} \left\vert \phi_i \right> .
 \label{Z}
\eeq
Here we denote the (approximate) energy eigenstate around a false vacuum as $\left\vert \phi_i \right>$.%
\footnote{One may think that the imaginary part of the vacuum energy is absent because $Z$ defined in \eq{Z} is real. This issue has been investigated in detail in Ref.~\cite{Andreassen:2016cff, Andreassen:2016cvx}. They discussed that the contour of path integral should be deformed along a steepest descent contour passing through the false vacuum. Fortunately, the resulting tunneling rate can still be calculated by \eq{Gamma_instanton}, which is the standard formula to calculate the tunneling rate by the instanton calculation. }
One may neglect the quantum fluctuation around the false vacuum and approximate the energy eigenstate by the eigenstate for the operator $\hat{\phi}$. In this case, $Z$ can be written as 
\beq
 Z \approx \int_{\phi(0) = \phi_i}^{\phi({\cal T}) = \phi_i} {\cal D} \phi \, e^{-S_E[\phi]}. 
 \label{Z2}
\eeq

The path integral can be approximated by the contribution from a saddle point, which is known as the instanton solution. One can also calculate the Gaussian integral for the perturbation around the instanton solution. The result is given by the well-known formula: 
\beq
 \Gamma_I \sim {\rm Im} K \, e^{-S_E[\phi_{\rm bounce}]}, 
 \label{Gamma_instanton}
\eeq
where $\phi_{\rm bounce}$ is the so-called bounce solution in an ``upside-down" potential, with $V\to-V$. Therefore, the decay rate can be calculated from the path integral with imaginary time ${\cal T}$. 

Strictly speaking, however, \eq{Gamma_instanton} is not a tunneling rate from the false vacuum energy eigenstate because the boundary condition for the path integral \eq{Z2} implies the transition between eigenstates for the operator $\hat{\phi}$. The difference between the energy eigenstate and the eigenstate for the operator $\hat{\phi}$ is negligible only if the zero-point fluctuation around the local minimum is much smaller than the typical scale of the potential. 

In quantum field theory, the number of effective degrees of freedom can be large and hence the quantum fluctuations can accidentally overcome the potential barrier. This accidental arrangement and subsequent barrier penetration was seen in the simulations of Ref.~\cite{Braden:2018tky}. Hence, to only focus on initial conditions that are eigenstates of the field operator, as the usual instanton approach does, is not guaranteed to be the most natural choice of boundary conditions. Therefore, we would like to utilize a formalism that can accommodate general initial conditions on the fluctuations for a more complete analysis of tunneling. In the next section, we analyze tunneling within the Wigner representation as it will allow us to systematically study these different possibilities.

\section{More General Formalism}\label{Real}

Since vacuum decay is a time-dependent process, it is natural to calculate it by using a real time formalism (or Schwinger-Keldysh formalism), which can describe the time evolution of observables. As we will see, this will allow us to more systematically identify initial conditions for the decay, rather than restricting to only those that are useful in the standard imaginary time analysis.

\begin{figure}[t] 
\centering
\includegraphics[width=8cm]{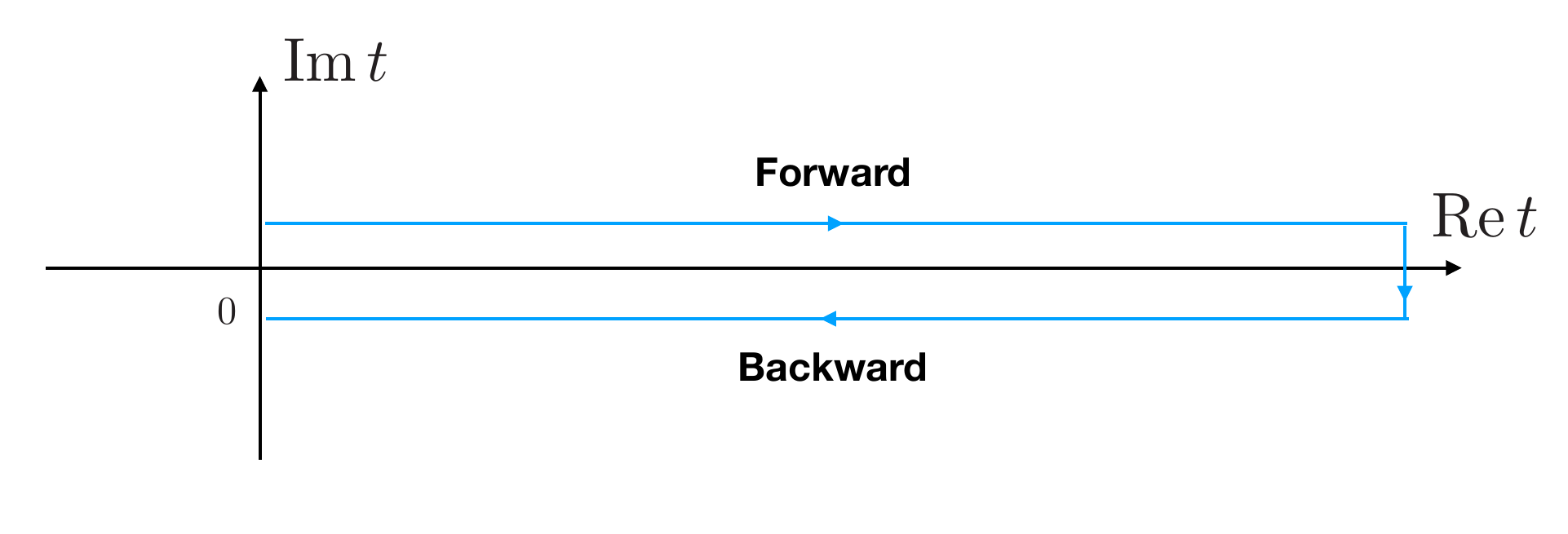} 
\caption{Integration contour for the time variable in the real time formalism.}
\label{fig}
\end{figure}

The time evolution of the expectation value of an observable $\hat{O}$ can in principle be calculated from 
\beq
 \la \hat{O}(t) \ra = \Tr \lkk \hat{\rho} \, T_{\rm time} \, e^{i \int \! H d t'}\, \hat{O} \,e^{-i \int\! H d t'} \rkk, 
\eeq
where $\hat{\rho}$ is an initial density operator and $T_{\rm time}$ is Schwinger's time-ordered operator. The operator $\hat{O}$ may be taken to be an order parameter of the phase transition. Instead, one may use an operator that gives zero around the false vacuum and nonzero around the true vacuum. This equation implies that the contour for the time integral is given by the one shown in Fig.~\ref{fig}. We can define two kinds of fields: forward ($\phi_f$) and backward ($\phi_b$) fields, depending on the direction of time evolution. It is convenient to then define 
\beq
 &&\phi_c = \frac12 \lmk \phi_f + \phi_b \rmk ,
 \\
 &&\pi_c = \frac12 \lmk \pi_f + \pi_b \rmk ,
 \\
 &&\phi_q = \frac12 \lmk \phi_f - \phi_b \rmk ,
 \label{phiq}
 \\
 &&\pi_q = \frac12 \lmk \pi_f - \pi_b \rmk ,
\eeq
where $\pi$ is the canonical conjugates of the field $\phi$. Here $\phi_c$ and $\pi_c$ are effectively classical fields, as we will explain shortly, while $\phi_q$ and $\pi_q$ are effectively quantum fluctuations. The path integral can then be written as 
\beq
&& \la \hat{O} (t_1) \ra = \int\! \int d \phi_{c,0} d \pi_{c,0} \,W_0 (\phi_{c,0}, \pi_{c,0}) 
 \nn
&&~~ \times \int {\cal D} \phi_c (t) {\cal D} \pi_c (t) 
 {\cal D} \phi_q (t) {\cal D} \pi_q (t) 
 \, O_W (\phi_c (t_1), \pi_c (t_1), t_1)
 \nn
&&~~ \times \exp \lkk i\! \int_{0}^{t_1} \! d t 
 \lkk 2 \phi_q \dot{\pi}_c - 2 \pi_q \dot{ \phi}_c 
 \right. \right. 
 \nn
&&~~ \left. \left.
+ H_W(\pi_c + \pi_q , \phi_c + \phi_q, t) 
 - H_W(\pi_c - \pi_q , \phi_c - \phi_q, t)
 \rkk \rkk,\,
 \nn
 \label{Ohat}
\eeq
where the Wigner function $W_0$ is defined as the Weyl transform of the density matrix $\hat{\rho}$ in the field representation. It is given by
\be
 W_0  (\phi_{c,0}, \pi_{c,0}) 
 = \int d \phi_q \,\rho ( \phi_{c,0} - \phi_q, \phi_{c,0} + \phi_q) e^{2i\pi_c \phi_q}.
\ee
Momentarily we focus on only one particular mode for notational simplicity, but will generalize shortly. The functions $H_W$ and $O_W$ are the Hamiltonian $H$ and observable $O$ in the Wigner representation, respectively. In particular, 
\be
 O_W (\phi_c, \pi_c) = \int d \phi_{q} d \pi_{q}\,
 O ( \phi_c - \phi_q, \pi_c + \pi_q ) \,
 e^{-2 i \phi_q \pi_q},
\ee
where $O(\phi, \pi)$ is the function obtained from the operator $\hat{O}$ by direct substitution $\hat{\phi} \to \phi$ and $\hat{\pi} \to \pi$. For related work, see Refs.~\cite{Kamenev:2009jj, Polkovnikov:2009ys}. 

\subsection{Classical Approximation}

Now we shall rewrite the above path integral by using some approximations. We will discuss the meaning and justification of these assumptions in the next section. 

If the quantum fluctuations are much smaller than the classical quantities, we can approximate $H_W$ as 
\beq
 &&H_W(\pi_c + \pi_q , \phi_c + \phi_q , t) 
 - H_W(\pi_c - \pi_q , \phi_c - \phi_q , t) 
 \nn
 &&\simeq 2 \pi_q \frac{\del H_W (\pi_c, \phi_c, t)}{\del \pi_c} 
 + 2 \phi_q \frac{\del H_W (\pi_c, \phi_c, t)}{\del \phi_c}. 
 \label{classical approximation}
\eeq
Then we can perform the integrations over $\phi_q$ and $\pi_q$, which give delta functions of the form
\beq
 \delta \!\lmk \frac{d \phi_c}{d t} - \frac{\del H_W}{\del \pi_c} \rmk 
 \delta \!\lmk \frac{d \pi_c}{d t} + \frac{\del H_W}{\del \phi_c} \rmk. 
\eeq
This means that indeed $\{\phi_c,\,\pi_c\}$ obey the classical equation of motion. The integrals over the fields are determined by the delta functions 
and the result is given by 
\beq
 \la \hat{O} \ra \simeq \int d \phi_{c,0} d \pi_{c,0} \,W_0 (\phi_{c,0}, \pi_{c,0}) 
 \left. O_W (\phi_c, \pi_c, t) \right\vert_{\rm classical}. 
 \nn
 \label{O_classical}
\eeq

If the initial quantum fluctuations are sufficiently small at the false vacuum, we can approximate the potential by a quadratic form; we will return to discuss under what conditions this approximation is valid. We denote the mass parameter as $m$ at the false vacuum, i.e., $V''(0)=m^2$. The ground state wave-function can then be approximated as
\beq
 \psi_0 \simeq \frac{1}{(\pi /\omega_k)^{1/4}} e^{- \omega_k| \phi_{c,0}|^2/2}, 
 \label{wavefunction}
\eeq
where $\omega_k = \sqrt{m^2 + k^2}$. The initial Wigner distribution $W_0$ can be then estimated by the one for a free field: 
\beq
 W_0 &=& \int d \phi_{q,0} \,
 \psi^* \lmk \phi_{c,0} + \phi_{q,0} \rmk 
 \psi \lmk \phi_{c,0} - \phi_{q,0} \rmk 
 e^{2 i \pi_{c,0} \phi_{q,0}} 
 \nn
 &\propto &  \exp\! \lkk - \lmk \omega_k |\phi_{c,0} (k)|^2 + \frac{1}{\omega_k} |\pi_{c,0}(k)|^2 \rmk \rkk\!. \,\,\,\,\,\,\,\,\,\,
 \label{initialdist0}
\eeq
This can be regarded as a probability distribution for the initial field values, with $\phi$ and $\pi$ treated as independent random variables. For quantum field theory in $d+1$ spacetime dimension, the full result for the initial Wigner distribution is approximated as
\beq
 W_0 \propto \exp\! \lkk - \int\! {d^{d}k\over(2\pi)^d} \!\lmk \omega_k |\phi_{c,0}(k)|^2 + \frac{1}{\omega_k} |\pi_{c,0}(k)|^2 \rmk \rkk\!, \,\,\,\,\,\,\,\,\,\,
 \label{initialdist}
\eeq
where we integrate over all $k$-modes.

We can calculate the tunneling rate by evolving the classical dynamics with an initial condition generated from \eq{initialdist}. Since the initial condition is generated randomly by the Wigner distribution, 
we should perform a large number of simulations to obtain a statistically reasonable result. 
The tunneling rate is therefore given by the statistical average of many realizations as done in Ref.~\cite{Braden:2018tky}. 
We note that the system has a translational symmetry and there is no strong correlation between two distant points. This implies that we can replace the ensemble average of many simulations by the spatial average of a single simulation with a large simulation box. 

The operator $\hat{O}$ should be taken such that it is nonzero around the true vacuum 
and is zero around the false vacuum. This can be realized by, e.g., taking $O = \theta (\phi - \phi_*)$ with $\phi_*$ being the field value at the other side of the potential barrier. Then we can count the number of nucleated bubbles per unit space as an exponential function of time. The tunneling rate is the coefficient of the time variable at the exponent~\cite{Braden:2018tky}.

\subsection{Relation to the Instanton Calculation}

The same result can be obtained by the saddle point approximation for \eq{Ohat}. In the classical limit, the path integral can be approximated by saddle points of the exponent of the integrand. Varying it with respect to $\phi_c$, $\phi_q$, $\pi_c$, and $\pi_q$, and eliminating $\pi_c$ and $\pi_q$, we obtain 
\beq
 &&2 \ddot{\phi}_c = - V' (\phi_c + \phi_q) - V' (\phi_c - \phi_q), 
 \label{EoM1}
 \\
 &&2 \ddot{\phi}_q = - V' (\phi_c + \phi_q) + V' (\phi_c - \phi_q). 
\eeq
One of the solutions to this equation is $\phi_q = 0$ with $\phi_c$ being the solution to the usual classical equation of motion, with initial conditions drawn from the initial Wigner distribution. This saddle point corresponds to \eq{O_classical}. We refer to this as the real time formalism from classical dynamics, seeded by non-trivial initial conditions that ultimately arise from a choice for the initial wave-function. Indeed we note that this process is absent if we were simply to assume trivial initial conditions $W_0  = \delta (\phi_{c,0}) \delta (\pi_{c,0})$, which would be the ``purely" classical behavior. 

Now we show that there is another, related, contribution to \eq{Ohat} that is non-zero even if we were to set $\phi_{c,0} = \pi_{c,0} =0$. Assuming $W_0  = \delta (\phi_{c,0}) \delta (\pi_{c,0})$, we rewrite \eq{Ohat} as 
\beq
&& \la \hat{O} \ra = 
\int {\cal D} \phi_f 
 {\cal D} \phi_b 
 \, O_W ((\phi_f +\phi_b) / 2, t_1)
 \nn
&&~~\,\,\,\,\,\,\,\,\,\,\,\,\,\,\,\,\,\,\,\,\,\,\,\,\,\,\,\, \times \exp \lkk  iS_f - iS_b \rkk,
\eeq
where we assume that $O_W$ is independent of $\pi_c$ and 
\beq
 S_{f,b} = i \int_{0}^{t_1} d t L_{f,b}, 
\eeq
are the actions for the forward and backward fields, respectively. This can be rewritten as 
\beq
&& \la \hat{O} \ra = 
 \int {\cal D} \phi 
 \, O_W ( \phi(t_1) , t_1)  \exp \lkk iS \rkk,
 \label{saddle2}
\eeq
by defining $t' \in (-t_1 , t_1)$ and $\phi (t) = \phi_f(t)$ for $t \in (0,t_1)$ and $\phi(t) = \phi_b(-t)$ for $t \in (-t_1, 0)$. This path integral can be calculated by the standard instanton method by deforming to imaginary time. Since we assume $\phi_{c,0} = \pi_{c,0} =0$ in this calculation, this saddle point corresponds to the transition from vanishingly initial classical fields. It also corresponds to vanishing initial quantum field for $\phi_q=0$, though it leaves the initial condition for the quantum field $\pi$ unspecified. This is identical to the one calculated by the instanton method discussed earlier in Section \ref{Imaginary}. It is therefore associated with going from a field eigenstate with Dirichlet boundary conditions and again returning, in imaginary time, to a field eigenstate with Dirichlet boundary conditions. It is the so-called bounce solution in imaginary time. Importantly, the difference from the saddle point solution corresponding to \eq{O_classical} is the initial condition (or the boundary condition at $t = 0$). 

Let us comment on how to rotate the time variable in the imaginary space. If we naively take $\phi_q = 0$ in \eq{Ohat}, the exponent vanishes. This is not consistent with \eq{saddle2}, where the action does not vanish and gives the Euclidean action in the imaginary time. This inconsistency comes from the naive analytic continuation of the time variable. We can use the epsilon prescription to specify a possible way to change the integration contour in the imaginary space. The Hamiltonian should include an imaginary mass term that specify the way to change the integration contour. Therefore the time variable for the Hamiltonian for $\phi_f$ should be rotated in the opposite way to the one for $\phi_b$. This is the reason that we obtain a nonzero exponent even if we take $\phi_q = 0$ in \eq{Ohat}.  Later we will comment on more general situations, which may occur in cosmology, where this rotation to imaginary time may be more problematic.

\subsection{Comparison}

In summary, there are two basic sets of initial conditions one may utilize to implement the saddle point for the path integral \eq{Ohat}. The first one is given by \eq{O_classical}, where the initial condition is given by some approximation to the initial wave-function and the time evolution is purely given by the classical equation of motion. The second one is given by the saddle point of \eq{saddle2}, where the initial condition is $\phi = 0$ and the time evolution is deformed into the complex plane to the imaginary time axis. 

At first sight it may seem surprising that the first should be associated with tunneling. But indeed tunneling can occur because of the non-trivial initial conditions can make for rare events to take place even within the framework of classical dynamics. This is the tunneling process that was calculated in Ref.~\cite{Braden:2018tky}. In this sense, this contribution is complementary to the instanton contribution, though in appropriate regimes that we will discuss, they can approximate each other quite well.

\section{Conditions for the classical approximation}\label{Condition}

In this section we discuss conditions to calculate a tunneling rate by \eq{O_classical} in the context of quantum field theory. We first note that the distinction between quantum and classical mechanics comes from the commutation relation for quantum operators. 
In particular, the commutation relation between creation and annihilation operators is given by 
\beq
 \hat{a}_i \hat{a}^\dagger_j -  \hat{a}_i^\dagger \hat{a}_j = \delta_{ij}. 
\eeq
However, the effect of the right-hand side is negligible when the occupation numbers $\la \hat{a}^\dagger_i \hat{a}_i \ra$ are large. We then expect that the high occupancy limit corresponds to the classical limit of quantum systems. This implies that the approximation \eq{classical approximation} is justified when the number of particles in the system is extremely large. 

In Ref.~\cite{Hertzberg:2016tal}, we have shown that the expectation values of quantum operators are approximated by a corresponding classical ensemble average over many classical micro-states, with initial conditions drawn from the initial quantum wave-function. Eq.~(\ref{O_classical}) is a mathematical expression of this statement. It can be understood as an extension of this discussion to the quantum regime, where the initial state is not a high occupancy state, but a (quasi) vacuum state with zero point fluctuations. Due to the possible production of bubbles, which arises due to rare accidental arrangements from the non-trivial initial conditions, the occupation number can be large enough to use the classical description. 

In this case, the approximation $\phi_q \ll \phi_c$ is satisfied, except for the initial condition, and we can evolve $\phi_c$ by the classical equation of motion. In the regime before the tunneling, $\phi_q \ll \phi_c$ may not be satisfied. However, we can still use \eq{O_classical} if the amplitude of fluctuations is small enough to neglect terms in the potential that are higher-order than quadratic. This is because the Wigner approximation is exact for the free-field theory. We will discuss situations in which the neglecting of these higher order terms may not be valid.

\subsection{Tension and Pressure}

\eq{O_classical} can describe the classical dynamics of the field after the bubble nucleation. This is different from the instanton method, where we need to connect the Lorentzian and Euclidean regimes to describes the dynamics of the bubble after nucleation. The tunneling process calculated by \eq{O_classical} can therefore describe the tunneling process itself as well as the dynamics of nucleated bubble after the nucleation. 

Since the nucleated bubble obeys the classical equations of motion, its behavior can be understood easily, particularly for the thin-wall case. The bubble wall tends to shrink to a point due to its tension while it tends to expand due to the pressure of the vacuum energy. As we evolve the field classically with an initial condition, a lot of small bubbles are nucleated, but most of them do not have enough pressure to overcome the tension of the wall. In order for the bubble to expand after the nucleation, the pressure of the vacuum energy should overcome the tension of the bubble. For a thin-wall bubble, 
this requires 
\beq
 A_{d-1} R^{d-1} \sigma \lesssim 
 V_d R^d \epsilon,
 \label{energy condition}
\eeq
where $R$ is the radius of the bubble, $\sigma$ is the tension of the wall, and $\epsilon$ is the difference of the vacuum energy. Here we define area and volume factors in the unit d-dimensional sphere: 
\beq
 A_{d-1} \equiv \frac{2\pi^{d/2}}{\Gamma(d/2)}, 
 \\
 V_d \equiv \frac{\pi^{d/2}}{\Gamma(d/2+1)}. 
\eeq
A similar type of inequality is expected to be satisfied for a thick-wall bubble.

\subsection{Occupation Number}

Now we examine under what conditions the occupation number of the quanta describing nucleated bubbles is much larger than unity. In this case the nucleated bubble is essentially coherent and can be treated within the framework of classical field theory. 

\begin{figure}[t] 
\centering
\includegraphics[width=8cm]{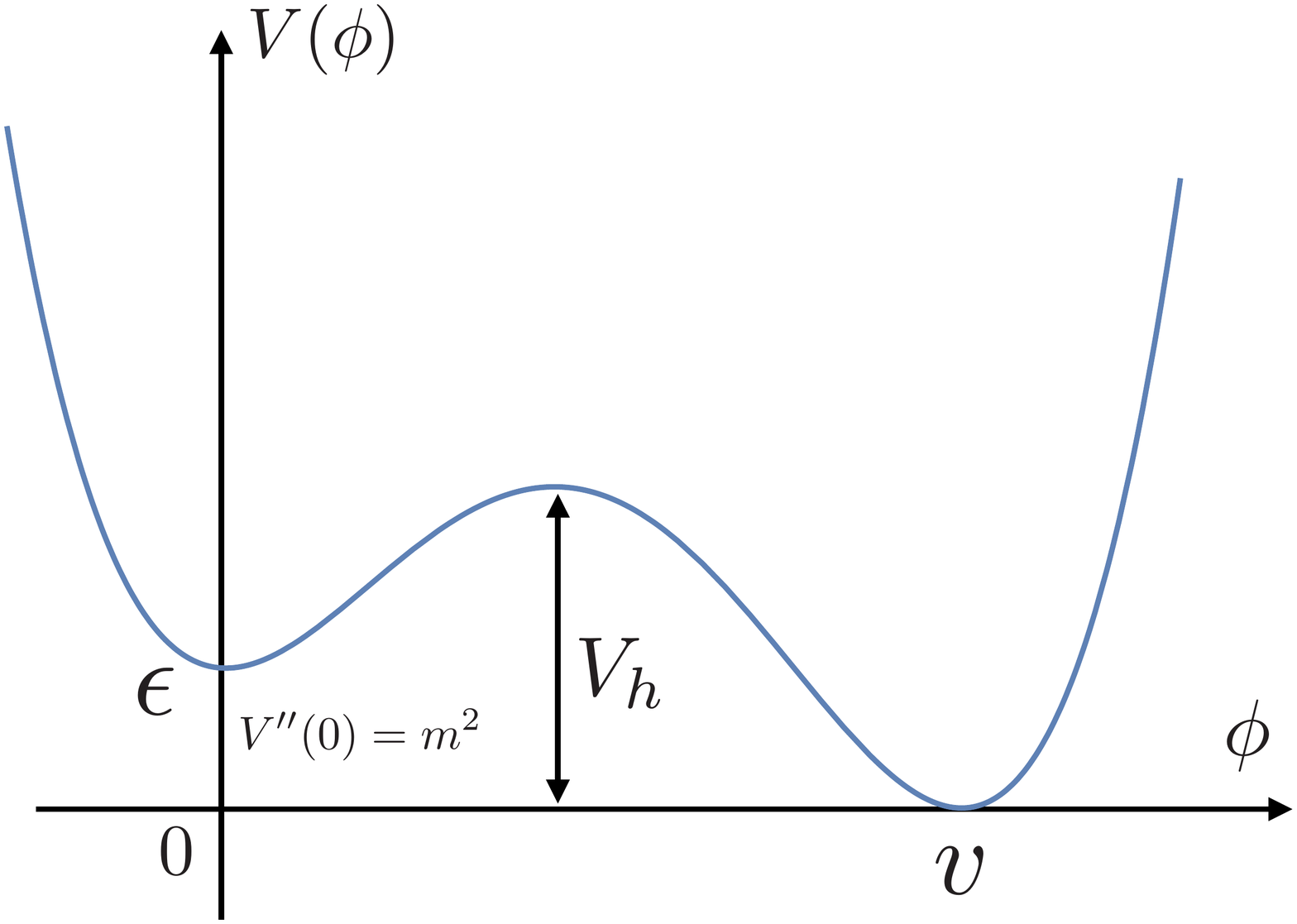} 
\caption{Schematic picture of a typical potential and parameters describing its shape.}
\label{fig2}
\end{figure}

We estimate the occupation number of nucleated bubble in two simple cases. First we consider the case where the scalar potential is described by typical values of curvature scale around vacua $m$, field value $v$, height of the potential barrier $V_h$, and the difference of the vacuum energy $\epsilon$ (see Fig.~\ref{fig2}). We note that $V_h$ must be smaller than of order $v^{2(d+1) / (d-1)}$ for $d>1$ because of the unitarity bound (e.g., in $3+1$ dimensions this is related to the familiar idea that the quartic coupling $\lambda\,\phi^4$ obeys $\lambda\lesssim 1$ to be in a weakly coupled regime). 

\subsubsection{Thin-Wall}

We assume $\epsilon \ll V_h$ and use the thin-wall approximation for now. In this case, the wall tension is given by 
\be
 \sigma = \int d \phi \,\sqrt{2\, V} \sim v \sqrt{V_h}. 
 \label{sigma}
\ee
Using \eq{energy condition}, we obtain a typical radius of the nucleated bubble as 
\be
 R_b \simeq d\, \frac{\sigma}{\epsilon}. 
 \label{R_b}
\ee

Let us first report on the Euclidean instanton action associated with the bubble, as it is the standard quantity to compute tunneling in the literature, as
\beq
 S_E 
&\simeq& \frac{2 R^d \sigma }{d+1} 
\simeq \frac{2 d^d}{d+1} \frac{\sigma^{d+1}}{\epsilon^d} 
\\
 &\sim& \lmk \frac{V_h}{\epsilon} \rmk^d \!\lmk \frac{v^{2(d+1) / (d-1)}}{V_h} \rmk^{(d-1)/2}. 
 \label{S_E}
\eeq

However, in order to justify the alternative real time tunneling from classical dynamics, we need to compute the bubble's occupancy number. We define it by the gradient energy of the bubble 
in the unit of $m$: 
\beq
 {\cal N} 
 &\equiv& \frac{1}{m} \int d^d x \lkk \frac12 \lmk \nabla \phi \rmk^2 
 \rkk_{\rm bubble}
 \\
 &\simeq& \frac{A_{d-1} \, R^{d-1} \sigma }{m}. 
\eeq
(if we were to reinstate factors of $\hbar$, the actual occupancy number would be this divided by $\hbar$). It is roughly given by
\be
 {\cal N}  \sim \lmk \frac{V_h}{\epsilon} \rmk^{d-1} \!\lmk \frac{v^{2(d+1) / (d-1)}}{V_h} \rmk^{d/2-1} 
 \! \lmk \frac{v}{m^{(d-1)/2}} \rmk^{2/(d-1)}\!, 
\label{Nthin}\ee
Note that every factor on the most right-hand side is larger than unity for weakly coupled field theories, so the occupation number can in fact be quite large ${\cal N}\gg 1$. (Note that for $d=1$, it simplifies to ${\cal N} \sim v^2 \, (V_h/(m^2v^2))^{1/2}$.) 

\subsubsection{Thick-Wall}

The above suggests that the occupation number can be as small as of order unity when the vacuum energy is not degenerate and the difference of the VEV is as small as $m$. This implies that the bubble is a thick-wall type and so we should re-examine the above analysis. Here the scalar self coupling could be as large as ${\cal O}(1)$. We check that the occupation number is larger than unity in this extreme case, too. We consider the following potential: 
\beq
 V(\phi) = U_0 \lkk \frac{1}{2} \lmk \frac{\phi}{\Lambda} \rmk^2 
 + \frac{\lambda_3}{3!} \lmk \frac{\phi}{\Lambda} \rmk^3 \rkk, 
\eeq
where $U_0$ and $\Lambda$ ($U_0 \lesssim \Lambda^{2(d+1)/(d-1)}$) are dimension-full parameters and $\lambda_3$ ($\lesssim {\cal O}(1)$) is a dimensionless constant. Since $\phi = 0$ is a false vacuum, it can tunnel to the other side of the potential hill. The tunneling action and the occupation number are given by 
\beq
&& S_E \simeq c_S  \lambda_3^{-2} A_d 
 \lmk \frac{\Lambda^{2(d+1)/(d-1)}}{U_0} \rmk^{(d-1)/2}\!,
 \label{S_E2}
 \\
&& {\cal N} \simeq c_{\cal N}  \lambda_3^{-2} A_{d-1} 
 \lmk \frac{\Lambda^{2(d+1)/(d-1)}}{U_0} \rmk^{(d-1)/2}\!,
 \label{N2}
\eeq
where the numerical constants are given by $c_S \simeq 4.1 \times 10$ and $c_{\cal N} \simeq 1.0 \times 10^2$ for $d =3$. Even in this extreme case, the tunneling action and the occupation number are larger than ${\cal O}(10/\lambda_3^2)$. This justifies that the nucleated bubble can be described classically (or a scalar condensate). 

\subsection{Single Particle Quantum Mechanics}

One may wonder if these arguments can extend to the problem of single particle tunneling in ordinary non-relativistic quantum mechanics. In this case there is obviously no such thing as a ``bubble" that can be formed. So there is no obvious sense in which there is any object at high occupancy.

Nevertheless, we can formally view this problem as quantum field theory in $0+1$ dimensions. So, for the sake of completeness, let us formally take the result in \eq{Nthin} and take $d\to0$. Then we formally obtain
\be
{\cal N}\sim {\epsilon\over m}.
\ee
Now we should note that in this case $m$, which is the (square root) of curvature of the potential in quantum field theory at the false vacuum, is just the characteristic frequency of oscillation of the particle $\omega_0$ around the meta-stable minimum in quantum mechanics. 

However, what is important is that in quantum mechanics, energy conservation tells us that the tunneling process requires that the particle tunnel to a point at the same potential energy as its starting values. Hence, $\epsilon$ here should be the potential height difference, so it is in fact just $\epsilon=0$. This implies ${\cal N}=0$. So there is no sense in which one is at high occupancy. This means that this procedure of sampling from some Gaussian approximation to the wave-function and using this to determine tunneling will typically fail in ordinary quantum mechanics. Conversely, it can be applicable in field theory in higher dimensions with bubbles of high occupancy, as we discussed above.

Furthermore, in single particle mechanics, with a confining potential, the system does not exhibit ergodicity. On the other hand, some ergodicity is exhibited in the field theory, providing the adequate time evolution for the development of bubbles over time as the system explores phase space. 

\section{Tunneling rate}\label{Tunneling}

Now we shall compare the tunneling rate (per unit volume) via the standard Euclidean instanton to the tunneling rate via classical dynamics with initial conditions drawn some approximation to the wave-function. The latter one can be estimated in the following way. 

Ideally the relevant initial fluctuations that ultimately lead to the formation of the bubble are sufficiently small that the potential can be approximated to be a quadratic form around a false vacuum; we will revisit this shortly. Then the initial distribution of fluctuations is given by the Wigner distribution of a free massive scalar field with mass $m$ around the false vacuum \eq{initialdist}. This distribution does not change much even if we allow the field to classically evolve in time. The tunneling rate can therefore be estimated by the probability that $\phi_{c} (k)$ and $\pi_{c} (k)$ are large enough to nucleate a classical bubble. A classical bubble that expands after the nucleation must satisfy the condition \eq{energy condition} and hence its radius must be larger than $R_b \sim \sigma/\epsilon$. 

Now in order to completely determine the probability for tunneling, one should perform a simulation of this non-linear system of classical equations of motion, with the appropriate initial conditions specified above. However, we can give an estimate of the probability of bubble formation by utilizing the initial wave-function's statistical distribution as a guide; we will return to this shortly. In order to form a bubble there are two conditions that need to be satisfied: (a) the field need's to be on the far side of the barrier and (b) the bubble needs to have sufficient energy to avoid collapse. Let us estimate these probabilities in turn

First, in order for the bubble to be on the other side of the barrier, we need that the field value in position space obeys $\phi_c\gtrsim v$. In the $k$-space representation this condition means that we need $\phi_c(k)>\phi_c^{(\rm th)} (k)$ where
\be
\phi_c^{(\rm th)} (k) = \left. \int d^d x\, e^{ikx} \phi(x) \right\vert_{\rm bubble} \sim R_b^d \, v, 
\ee
for a bubble of radius $R_n$ to be nucleated. The probability $P_a$ that $\phi_c(k)$ exceeds this threshold can be estimated from the Wigner distribution as
\be
 - \ln P_a \sim \int\! {d^d k\over(2\pi)^d}\, \omega_k |\phi_c^{(\rm th)}(k)|^2 
 \sim  \omega_b \,v^2\, R_b^d, 
 \label{P1}
\ee
where $\omega_b = \sqrt{m^2 + k_b^2} \sim \sqrt{m^2 + R_b^{-2}}$ is some characteristic frequency, associated with the bubble associated with a characteristic wavenumber $k_b \sim 1/R_b$. 

However, this is not a sufficient condition for nucleation, because if the bubble appears on the other side of the barrier with arbitrarily low energy then it can collapse. Suppose that a small bubble with radius $R_{\rm ini}$ and kinetic energy $R_{\rm ini}^d \pi_c^2$ forms due to fluctuations. The kinetic energy must be larger than the energy of the bubble with radius $R_b$ so that the bubble can expand after the nucleation. This means we need $\pi_c > \pi_c^{(\rm th)} $ where
\be
 R_{\rm ini}^d \lmk \pi_c^{(\rm th)} \rmk^2 \sim R_b^d \,\epsilon. 
\ee
Note that this $\pi_c^{(\rm th)}$ is the conjugate momentum in position space. It can be written in terms of $\pi(k)$ in the momentum space as $\pi_c^{(\rm th)} = 1/(2\pi)^d \int d^d k e^{-ikx} \pi_c(k) \sim \pi_c^{(\rm th)} (k \sim 1/R_{\rm ini}) R_{\rm ini}^{-d}$. The probability $P_b$ to have sufficient energy can then again be estimated from the Wigner distribution as
\be
 - \ln P_b \sim \int {d^d k\over(2\pi)^d} \frac{1}{\omega_k} |\pi_c^{(\rm th)}(k)|^2
 \sim    \frac{\epsilon}{\omega_b} R_b^d.
 \label{P3}
\ee

\subsection{Tunneling Rate}

According to the Wigner representation, the variables $\phi_c$ and $\pi_c$ are taken as independent random variables. This says that the probability that both (a) and (b) occur is the product $P_a\, P_b$. This allows us to estimate the tunneling rate (per unit volume) within this real time formalism as
\be
 \Gamma_R \sim c_R\,e^{-\gamma_R}, 
\ee
with
\be
\gamma_R = a\,\omega_b \,v^2\, R_b^d+b\, \frac{\epsilon}{\omega_b} R_b^d
\label{gamRapprox}\ee
where $a$ and $b$ are $\mathcal{O}(1)$ prefactors.

We can compare this to the usual result for tunneling using the Euclidean imaginary time formalism given earlier in \eq{Gamma_instanton} as \be
\Gamma_I\sim c_I\,e^{-\gamma_I},
\ee
with
\be
 \gamma_I = S_E \sim \epsilon \,R_b^{d+1}.
 \ee
The instanton rate is calculated for the thin-wall case, but the result is not qualitatively different for the thick-wall case once we identify $\epsilon$ as the difference of the vacuum energy between the false vacuum and the tunneling point. Since this involves an extra factor of $R_b$ compared to the scaling in $\gamma_R$, we need an estimate for the bubble radius, which is roughly
\beq
 R_b 
 &\sim& 
 {v\sqrt{V_h}\over\epsilon},
\eeq
This allows us to make the estimate
\be
\gamma_I\sim v\sqrt{V_h} \, R_b^d,
\label{gamApprox}\ee
for the instanton tunneling exponent. In this final expression we have still kept a factor of $R_b^d$ for convenience, since this is a common factor that appears in \eq{gamRapprox} also.

\subsection{Examples}

We now use the above results to compare the tunneling rates that we have estimated in these different formalisms.

\subsubsection{Weakly Broken $\mathbb{Z}_2$ Symmetry}

Let us consider a potential of the form 
\be
V(\phi)=\lambda(\phi^2-v^2)^2+\delta V(\phi), 
\ee
where $\delta V(\phi)$ is a term that weakly breaks the $\mathbb{Z}_2$ symmetry.  This potential is similar to the kind of potential shown in Fig.~\ref{fig2} with $V_h \sim m^4/\lambda$. In this case the bubble thickness is approximately set by the Compton wavelength as $\lambda_C\sim 1/m$. However the bubble radius is at least this large, i.e., $m\,R_b\gtrsim 1$. This ensures the  frequency $\omega_b$ can be approximated by the mass $\omega_b\simeq m$. By noting that $\epsilon$ is bounded to be of the order of or much smaller than $V\sim v^2\,m^2$, we can conclude that the probability $P_a\lesssim P_b$. Hence the rate $\gamma_R$ is approximated as
\be
\gamma_R\sim m\,v^2\,R_b^d.
\ee
Then from \eq{gamApprox}, with $V_h/m\sim m\, v^2$, we have $\gamma_I\sim \gamma_R$.

\subsubsection{SM Higgs}

As another example, let us consider the Higgs potential in the minimal SM. Upon RG running of the Higgs self-coupling $\lambda$, the top mass, and other couplings, one finds that the Higgs potential turns over and then goes negative. This happens at around $v\sim 10^{11} \GeV$, or so. In this example, the potential is dominated by the quartic term near the tunneling point. In this case, we can use the above formula by taking $\epsilon \to V_h \sim \lambda\, v^4$. The bubble radius is now of order or larger than 
\be
R_b\sim{1\over\sqrt{\lambda}\,v}\sim 10^{-10}\,\mbox{GeV}^{-1},
\ee
(using $\lambda\sim 0.01$ in this regime). This radius is much much smaller than the Compton wavelength of the Higgs which is $m^{-1}\sim 10^{-2}\,\mbox{GeV}^{-1}$. Hence now we are in a regime in which $\omega_b\sim 1/R_b$. In this regime, both $P_a$ and $P_b$ are comparable, and they both give 
\be
\gamma_R\sim {1\over\lambda},
\ee
(we naturally focus here on the physical case of 3+1 dimensions). This is comparable to the instanton rate $\gamma_I\sim 1/\lambda$, so we again have $\gamma_I\sim \gamma_R$.

We note that in this case with $R_b\ll m$, giving $\omega_b\gg m$, and probing deep into the quartic term in the potential, it was not guaranteed that the Gaussian approximation based on the free theory would suffice. However, parametrically it is of the right order.

\subsubsection{Flat Hill-Top}

Suppose the hill-top is very flat, moreso than it appears in Fig.~\ref{fig2}. To be clear, let us imagine that it is so flat that $V_h\ll m^2 v^2$, which would be the naive value based on dimensional analysis.  Such a potential is perhaps unusual from the microscopic point of view, but it is allowed in principle. In this case the instanton gives an exponent (normalized to bubble volume) that is linear in the barrier width $\gamma_I/R_b^d\propto v$. On the other hand, if we turn to the real time formalism we obtain different estimates. From \eq{P3} the contribution from the kinetic energy effect gives $\gamma_R/R_b^d\propto v^0$, which is too small. On the other hand, from \eq{P1} the contribution from the need to be on the other side of the barrier gives $\gamma_R/R_b^d\propto v^2$, which is too large. 

In this case, the Gaussian approximation for the initial wave-function is not accurate, since it assume that the fields mass is $m$, but for such a potential, the effective mass in the barrier is smaller. Instead we need to alter our simple estimates. We need to essentially replace the frequency of the bubble by some appropriate effective mass, from the effective curvature of the potential, namely $m_{\rm eff}\sim\sqrt{V_h}/v$. This leads to
\be
-\ln P_a\to m_{\rm eff}\,v^2\,R_b^d\sim v\sqrt{V_h}\,R_b^d,
\ee
which is indeed of the order of the instanton rate. 

On the other hand if we persist with the original Wigner distribution, we believe that it is plausible that a simulation can arrive at roughly the correct tunneling rate anyhow. This is because even though the initial distribution is not an accurate representation of the false vacuum eigenstate, these initial conditions may be partially washed away in the simulation, leading to the appropriate rate.

\section{Discussion}\label{Discussion}

We have used a general Wigner representation to establish two formulations of tunneling with slightly different boundary conditions and dramatically different dynamics: in addition to the usual formulation of the imaginary time saddle point contribution to the decay amplitude with Dirichlet boundary conditions on the field, there is another real time formulation based on classical dynamics with initial conditions set by some estimate for the initial wave-function. While the former one is the familiar one from the instanton action, the latter one is an ensemble average of classical field theory dynamics seeded by quantum zero point fluctuations. We note that this ensemble average can be practically realized by a spatial average of a single simulation by appealing to a form of ergodic theorem. 
Since we use the Wigner approximation for the initial wavefunction 
in the real-time approach, we do not expect that the resulting rate is exactly the same as the 
one calculated in the imaginary-time approach. 
However, we have checked that the exponent in the tunneling rate is parametrically the same in both approaches in several examples.

\subsection{Classicality}

In order to justify the classicality of the field in this latter approach, the quantum fluctuations have to organize into a bubble and the occupation number has to be much larger than unity. This can be realized only if the degrees of freedom in the system are large enough, as is possible in quantum field theory, as it is for the nucleated bubbles. We note however that much of the universe would remain at low occupancy, so it is not entirely guaranteed that the classical dynamics is extremely accurate, but perhaps only roughly accurate. Furthermore, this approach is ordinarily not valid in single particle quantum mechanics as the notion of high occupancy there does not seem to be valid.

One may wonder if the tunneling rate depends sensitively on the initial fluctuations. This is actually the case when the number of degrees of freedom in the system is not much larger than of order unity. However, we are interested in the tunneling process in quantum field theory, where the number of relevant degrees of freedom can be quite large. In this case, all relevant modes may interact with each other somewhat chaotically and the distribution will be randomized after an ergodic time. So one expects dynamical evolution to wash away some features of the initial condition. 
(We may have to wait for a time scale longer than the time scale of oscillation around a false vacuum so that some features of the initial condition are washed away. This condition is similar to $T/ T_{\rm slosh} \to \infty$ for the ``direct method" that was introduced in Refs.~\cite{Andreassen:2016cff, Andreassen:2016cvx}.)

However, our simple estimates for the tunneling rates did involve a dependence on the mass of the field defined around the initial false vacuum, as it affects the initial Gaussian approximation to the wave-function. So these simple estimates involve some sensitivity to initial conditions, especially in the case of potentials with extreme features. But in the case in which the bubble has characteristic wavenumbers $k$-values ($k\lesssim m$), we are not sensitive to the UV behavior of the initial conditions. Furthermore, more general estimates could be made in more extreme situations also.

\subsection{Applications}

As an application of these results, suppose there is an AdS vacuum between two dS vacua. The tunneling rate from a dS vacuum to the other dS vacuum cannot be calculated by using the standard instanton method because there is no instanton solution. However, the transition rate must be nonzero because anything can happen in quantum theory according to the path integral expression~\cite{Brown:2011ry}. In fact, the ``classical tunneling'' discussed in this paper is expected to give a nonzero transition rate. This is the only practical way we are aware of to calculate the transition rate in such a case. This transition process is complementary to the standard instanton tunneling process. In this sense, the result gives a lower bound on the tunneling rate. 

As another application, consider a dynamical setting, such as during preheating after inflation. In this case a field may exhibit a strongly time dependent effective potential from its interactions with the inflaton or the metric etc. If such a field is also trapped in a type of false vacuum then it may be highly non-trivial to implement the standard instanton tunneling procedure, as this requires deforming the contour to the imaginary time axis. If there is (quasi) periodic behavior in the time domain it will re-organize into growing exponential behavior in imaginary time, which may be an obstruction to an efficient implementation of the Euclidean instanton analysis. Furthermore, if there is some form of non-analytic structure to the time dependence, such as from a step-like time behavior, then this may be an obstruction to deforming the contour. In these cases it may be more intuitive and more practical to perform a real time analysis. 

\vspace{0.7cm}

\section*{Acknowledgments}

We would like to thank Jonathan Braden, Alexander Vilenkin, and Mohammad Hossein Namjoo for useful discussions. MPH is supported by National Science Foundation grant PHY-1720332 and a JSPS fellowship. 

\bibliography{reference}

\end{document}